\documentclass[12pt]{article}

\usepackage{authblk}

\usepackage{amsmath,amssymb}
\usepackage[version=3]{mhchem}     
\usepackage{subfig}               
\usepackage{graphicx}             
\usepackage{hyperref}             

\usepackage{booktabs}    
\usepackage{subcaption}  
\usepackage{siunitx}     

\title{Electron - Phonon Temperature Inversion in Nanostructures under Pulsed Photoexcitation}

\author[1]{Qian Ye}
\author[2,3]{Stephen K. Sanders}
\author[1,4]{Andrea Schirato}
\author[2,5]{Alessandro Alabastri\thanks{alessandro.alabastri@rice.edu}}

\affil[1]{Department of Physics and Astronomy, Rice University, Houston, TX, 77005, USA}
\affil[2]{Department of Electrical and Computer Engineering, Rice University, Houston, TX, 77005, USA}
\affil[3]{Oak Ridge Associated Universities, U.S. Army DEVCOM Army Research Laboratory-South, Houston, TX, USA}
\affil[4]{Department of Physics, Politecnico di Milano, Milano, Italy}
\affil[5]{Smalley-Curl Institute, Rice University, Houston, TX, 77005, USA}

\date{\today}

\begin{document}
\maketitle

\begin{abstract}
Photoexcitation of metallic nanostructures with short optical pulses can drive non-thermal electronic states, which, upon decay,  lead to elevated electronic temperatures (\(T_e \gtrapprox 1000\,\mathrm{K}\)) eventually equilibrating with the lattice (\(T_p\)) through electron-phonon scattering. Here, we show that, in spatially extended nanostructures, the lattice temperature can locally exceed that of the electrons, a seemingly counterintuitive transient effect termed hereafter ``temperature inversion'' (\(T_p > T_e\)). This phenomenon, fundamentally due to inhomogeneous absorption patterns and absent in smaller particles, emerges from a complex spatio-temporal interplay,  between the electron-phonon coupling and competing electronic thermal diffusion. By combining rigorous three-dimensional (3D) finite-element-method-based simulations with practical reduced zero-dimensional (0D) analytical models, we identify the electron-phonon coupling coefficient (\(G_{e-p}\)) as the critical parameter governing this behavior. An optimal \(G_{e-p}\) range allows the inversion, whereas a weak or overly strong coupling suppresses it. Among common plasmonic metals, platinum (Pt) exhibits the most pronounced and long-lived inversion, while gold (Au) and silver (Ag) show no significant inversion. Moreover, the close agreement between the 0D and 3D results, once an appropriate characteristic length is selected, highlights that the essential physics governing the inversion can be captured without full spatial complexity. These results provide insights for optimizing nanoscale energy transfer and hot-carrier-driven processes, guiding the strategic design of materials, geometries, and excitation conditions for enhanced ultrafast photothermal control.
\end{abstract}

\section{Introduction}
Understanding the ultrafast photothermal response of metallic nanostructures following 
pico- or femtosecond optical excitation is crucial to guide a wide range of applications in nanoscale energy conversion,\cite{NSA2016} photothermal therapies,\cite{BQ2013,AWE2019} plasmonic device engineering,\cite{LLH2007,AP2010,C2014} and hot-carrier science \cite{BHN2015}. 
At the core of these processes lie the intricate nanoscale interactions between the incoming photons, the conduction electrons in the metal, and its lattice (phonons). 
Upon laser illumination and photo-absorption, a collective oscillation of charges is induced in the nanostructure, known as Localized Surface Plasmon (LSP) \cite{M2007}. Over time, the LSP dephases quickly and, by decaying non-radiatively, it releases its excess energy to highly nonequilibrium (hot) electrons \cite{MLK2014,BKW2017}. In their higher-energy states, these hot carriers can reach elevated temperatures (up to thousands of Kelvin) before transferring energy to the phonons via electron-phonon scattering events towards equilibrium \cite{SMC2023,CN2022}. 
When considering relatively large nanoparticles, or nanostructures featuring peculiar morphologies \cite{HLC2011,MSB2022} (e.g., with spikes, tips, or edges), the interplay between electron-phonon coupling, heat diffusion for both the electronic and the phononic temperature, and geometric photo-absorption localization may lead to complex, time-dependent energy exchange mechanisms \cite{SCZ2022, Rudenko_AdvOptMat_2018, Sivan_ACSP_2020, Rudenko_PRB_2021, Bryche_ACSP_2023}. 
An  intriguing manifestation of the interplay of such processes is a transient, local ``temperature inversion'', in which the phonon temperature ($T_p$) locally exceeds the electron temperature ($T_e$), challenging the intuitive expectation that electrons, responding more quickly, should remain hotter at early timescales. 

In this work, we perform a systematic numerical investigation of the dynamic photothermal response of metallic nanostructures under pico- and femtosecond pulsed excitation. 
We combine reduced 0D analytical models with 3D finite-element-method (FEM)-based simulations to elucidate fundamental scaling behaviors, material-specific trends, and geometric effects. 
The 0D approach allows for prompt parametric studies, highlighting the pivotal role of the electron-phonon coupling coefficient ($G_{e-p}$) and the temperature-dependent parameters. 
On the other hand, our 3D models provide space-resolved insights into how geometry, diffusion, and nonuniform energy dissipation govern the local photothermal dynamics in extended nanostructures. 
By comparing 0D and 3D results, we show that the essential physics of the temperature inversion mechanism analyzed here can be captured by a spatially uniform description if an appropriate characteristic length mimicking electronic heat diffusion is selected, allowing us to test the impact of each relevant parameter efficiently before resorting to computationally intensive 3D simulations.

Our results reveal that platinum (Pt) exhibits pronounced and long-lived temperature inversions, as opposed to other metals (e.g., Au, Ag), which show negligible inversion, mostly owing to weaker electron-phonon coupling or slower electron relaxation. 
Intermediate responses (Cu, Al, Ni) align with their coupling strengths. 
We further demonstrate how the variation of the exciting pulse fluence and pulse duration affect both the temporal window and the magnitude of such inversion. 
Notably, at sufficiently high fluences (e.g., $F > 100\,\mathrm{mJ/cm^2}$), inversion vanishes entirely, illustrating a nonlinear dependence on the excitation conditions. 
Intermediate fluences and pulse durations yield optimal inversion due to the temperature dependence of electron thermal conductivity ($k_e$) and heat capacity ($C_e$), which regulate the relaxation timescales and excess energy exchange.

By providing a detailed comparative analysis of electron-lattice temperature inversions for various materials, geometries, and excitation conditions, this study contributes to the fundamental understanding of the ultrafast photothermal responses in plasmonic nanostructures. 
These results provide insights for designing materials and devices and optimizing electronic energy harvesting, controlled photothermal heating, and nanoscale heat management. 
The combination of 0D and 3D approaches forms a complementary framework, paving the way for predictive design strategies in hot-carrier-driven applications.

\section{Methods}

\subsection{Three-Dimensional Finite-Element Modeling}
To capture geometric effects and space-resolved energy exchange mechanisms between incoming photons, electrons and the metal lattice, we use the inhomogeneous, local formulations \cite{SMC2023,SCZ2022} of the well-established three-temperature model (3TM) \cite{SVA1994}. 
In essence, the 3TM  is a rate-equation model detailing the photoexcitation and relaxation of plasmonic nanostructures upon illumination with ultrashort laser pulses in terms of three energetic degrees of freedom: $N_{NT}$, the excess energy stored in a nonthermal fraction of the electronic population, whose distribution deviates substantially from a Fermi-Dirac; $T_e$, the temperature of hot carriers closer to the Fermi level, whose energy distribution can be described by a nonequilibrium Fermi-Dirac distribution; and $T_p$, the metal phononic temperature.
Further details on our modeling approach can be found elsewhere \cite{SMC2023,ZPK2015}.

We implement this model in 3D FEM simulations using COMSOL Multiphysics to describe both electromagnetic and energy transfer mechanisms with spatial and temporal resolution. 
The systems under study (refer to the main text) consist of spherical nanoparticles of different sizes and a metallic bowtie-shaped nanostructure embedded in a dielectric environment.
In formulas, our inhomogeneous 3TM reads as follows, as a function of time $t$ and the position across the metal nanoparticle $\mathbf{r}$: 

\begin{align}
N_{\rm NT}(\mathbf{r},t) &= \int_{-\infty}^{t} e^{-(a+b)(t-t')} Q_{\mathrm{abs}}(\mathbf{r},t')\,dt'.
\label{eq3}\\[6pt]
C_e(T_e)\frac{\partial T_e}{\partial t} &= \nabla \cdot [k_e(T_e)\nabla T_e] - G_{e-p}(T_e-T_p) + a\,N_{NT}(\mathbf{r},t), \label{eq1}\\[6pt]
C_p(T_p)\frac{\partial T_p}{\partial t} &= \nabla \cdot [k_p(T_p)\nabla T_p] + G_{e-p}(T_e-T_p) + b\,N_{NT}(\mathbf{r},t),
\label{eq2}
\end{align}
Here, the quantities $C_e(T_e)$ and $C_p(T_p)$ represent the temperature-dependent heat capacities of the electron and phonon subsystems, $k_e(T_e) = k_p T_e / T_p$ (following \cite{KSI1998}), \(k_p(T_p)\) are the corresponding thermal conductivities, $G_{e-p}$ is the electron-phonon coupling coefficient.
According to the metal considered, values for $C_e(T_e)$ and $G_{e-p}$ were taken from \cite{LZC2008},

The electromagnetic power density of the absorbed pulse is $Q_{\mathrm{abs}}$ and can be expressed as \cite{SCZ2022,ZPK2015}:
\begin{equation}
Q_{\mathrm{abs}}(\mathbf{r}, t) = F \frac{Q_{\mathrm{rh}}(\mathbf{r})}{I_0} \exp\left[-\frac{4 \ln(2)(t - t_0)^2}{\tau_{pulse}^2}\right],
\end{equation}
where $F$ is the pulse fluence, $Q_{\mathrm{rh}}(\mathbf{r})$ represents the spatial absorption profile (related to the material's absorption properties and geometry), and $I_0$ is the incident intensity. The Gaussian temporal profile is centered at $t_0$ and characterized by its full width at half maximum (FWHM), \(\tau_{pulse}\), which defines the pulse’s temporal duration.

The absorbed power density $Q_{\mathrm{abs}}$ generates a nonthermal electron energy density $N_{\mathrm{NT}}$, as described in Eq.~\ref{eq3}. The terms $a$ and $b$ characterize the energy transfer rates of nonthermal electrons:  $a$ is the rate at which nonthermal electrons transfer energy to the thermalized electron system, influenced by the electron-electron interactions and the Fermi energy; $ b $ represents the rate at which nonthermal electrons couple to the phonon system \cite{SMC2023}.

Together, these rates describe the relaxation pathways of the nonthermal electrons, driving the energy redistribution between the electronic and lattice subsystems.

\begin{figure}
    \centering
    \includegraphics[width=1\textwidth]{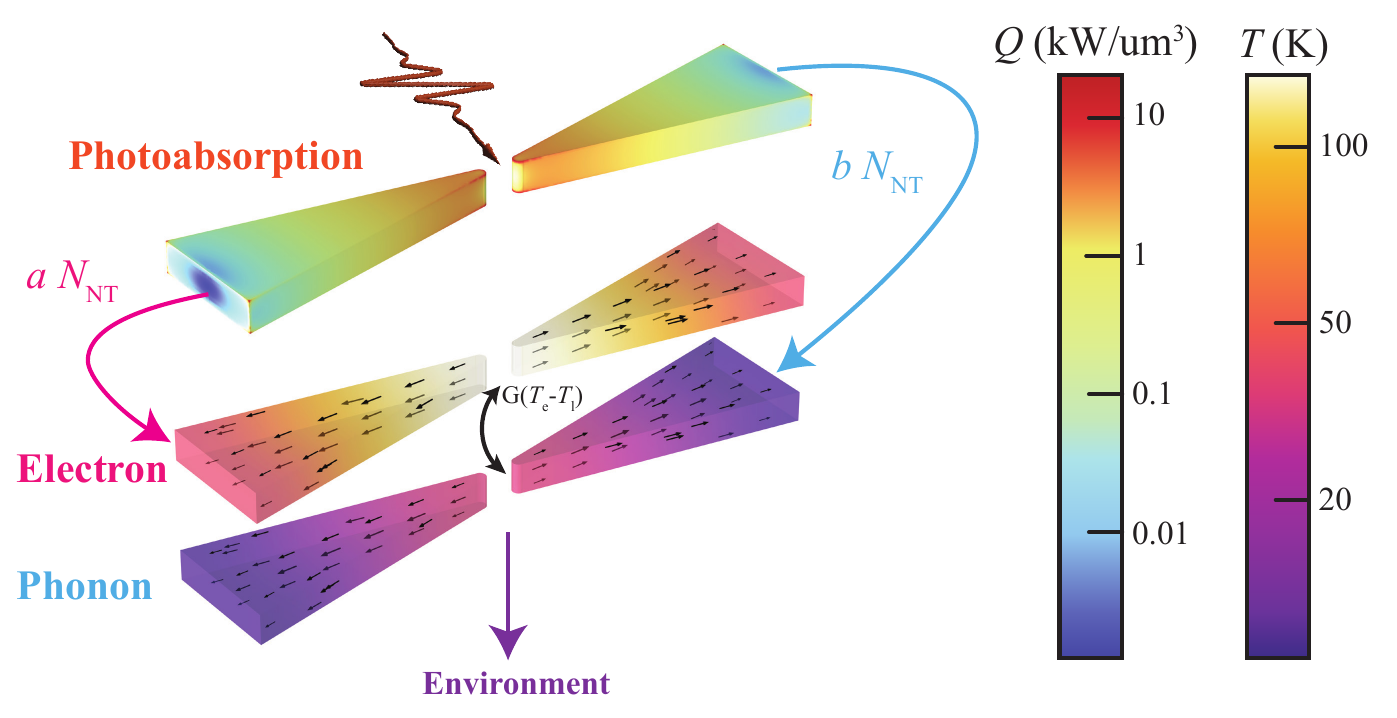}
    \caption{Schematic snapshot (at \( t \approx 6.1\,\mathrm{ps} \)) of the transient energy flow and thermal coupling in a metallic nanostructure under femtosecond laser illumination (600 nm wavelength, 75 \(\mu\mathrm{J}/\mathrm{cm}^2\) fluence, 2 ps duration). The shown time frame corresponds to the peak of the electron-phonon temperature difference (\( T_e - T_p \)) and is near the illumination pulse maximum. The top bowtie-shaped 3D structure represents the laser-induced heat source (dissipation rate), the middle layer shows the electron temperature distribution, and the bottom layer depicts the phonon temperature field. Heat fluxes in the electron and phonon layers are represented by black arrow fields. The terms \( aN_{NT} \),  \( bN_{NT} \) and \(G(T_e - T_p\)) correspond to energy transfer rates from nonthermal electrons to thermalized electrons and phonons and between thermalized electrons and phonons, respectively. Color bars on the right provide quantitative scales for the dissipation rate \( Q \) (in kW/$\mu$m$^3$) and the temperature  increase $\Delta T$ (in K) relative to the ambient temperature of 293.15 K.}
    \label{fig1}
\end{figure}

As exemplary geometries, we consider a bowtie structure composed of isosceles triangular prisms (height 45 nm, base width 20 nm, thickness 5 nm) separated by a 5 nm gap and nanospheres of varying radii (from 5 nm to 50 nm). 
The nanostructures are surrounded by a spherical environment (radius 250 nm), with perfectly matched layers and infinite element layers occupying the outer third of this radius to minimize artificial reflections. The mesh for spherical nanoparticles is designed based on the sphere’s radius ($R$), with the maximum mesh element size set to $R/10$ and the minimum mesh element size set to $R/30$. This ensures an adequate resolution of spatial features, particularly for capturing dissipation gradients and temperature inversion phenomena across different sphere sizes. For the bowtie-shaped nanostructure, the mesh is refined to a maximum element size of approximately 0.34 nm in the bowtie region, ensuring that steep temperature gradients are well resolved.
We employ the generalized-alpha method for temporal discretization, providing stable, nonoscillatory solutions suitable for capturing smooth transient processes at ultrafast timescales. 

Adaptive meshing and strict convergence criteria ensure that the solutions to Eqs.~(\ref{eq1}) and (\ref{eq2}) are numerically stable and physically meaningful, offering comprehensive spatial and temporal insights into the electron-phonon interactions.

\subsection{Zero-Dimensional Analytical Framework}
We consider a 0D 3TM to capture the essential electron-phonon energy exchange without spatial complexity.
Assuming uniform electron ($T_e$) and phonon ($T_p$) temperatures, the 0D-3TM we have formulated reads:
\begin{align}
C_e(T_e)\frac{\mathrm{d}T_e}{\mathrm{d}t} &= Q(t) - G_{e-p}(T_e-T_p) - \frac{k_e}{(bH)^{2}}T_e, \label{eq4}\\[6pt]
C_p(T_p)\frac{\mathrm{d}T_p}{\mathrm{d}t} &= G_{e-p}(T_e-T_p) - \frac{k_p}{(bH)^{2}}T_p. \label{eq5}
\end{align}
\noindent
The terms $\kappa_e=k_e / (b H)^2$ and $ \kappa_p=k_p / (b H)^2$ represent simplified heat-loss mechanisms that account for spatial heat diffusion using, replacing the heat flux divergence with the square of a characteristic length $bH$ serving as fitting parameter. Here, $H$ is the bowtie's height, and $b = 0.45$ is a fitting parameter. This choice ensures that the 0D model mimics the diffusion-like losses present in the full 3D model while maintaining computational efficiency. By setting $b \approx 0.45$, the 0D model effectively captures the heat-loss behavior at the bowtie scale, bridging the simplified approach with the detailed 3D results.

Unlike the 3D model, the 0D approach does not explicitly account for nonthermal electrons. Instead, the source term, previously represented by $a N_{NT}(\mathbf{r}, t)$ in 3D, is modeled as a time-dependent Gaussian source: 
\begin{equation}
I(t) = I_0 \exp\left(-\frac{4 \ln(2) (t - t_0)^2}{\tau_{\text{pulse}}^2}\right),
\end{equation}
where $I_0$ is the peak intensity, $t_0$ is the pulse center, and 
\(\tau_{\text{pulse}}\) is the full width at half maximum (FWHM) of the pulse. Throughout, the electron-phonon temperature difference, \(\Delta T_{e-p} = T_e - T_p\), remains our key metric for identifying temperature inversion.
Looking for general and insightful trends, we use the parameters $C_e, C_p,$ and $ G_{e-p}$ to explore how material properties and pulse conditions affect the temperature inversion. 
The computational efficiency of the 0D model enables broad parameter scans, guiding us toward promising regimes before resorting to more computationally intensive 3D simulations.

\section{Results and Discussions}
\subsection{Small Particle Case}

We begin by examining the ultrafast photothermal dynamics in a three-dimensional (3D) geometry composed of a single small spherical particle of radius 5 nm, aiming at investigating the role of different materials over electron and phonon temperature dynamics. 

\begin{figure}[h]
    \centering
    \includegraphics[width=1\textwidth]{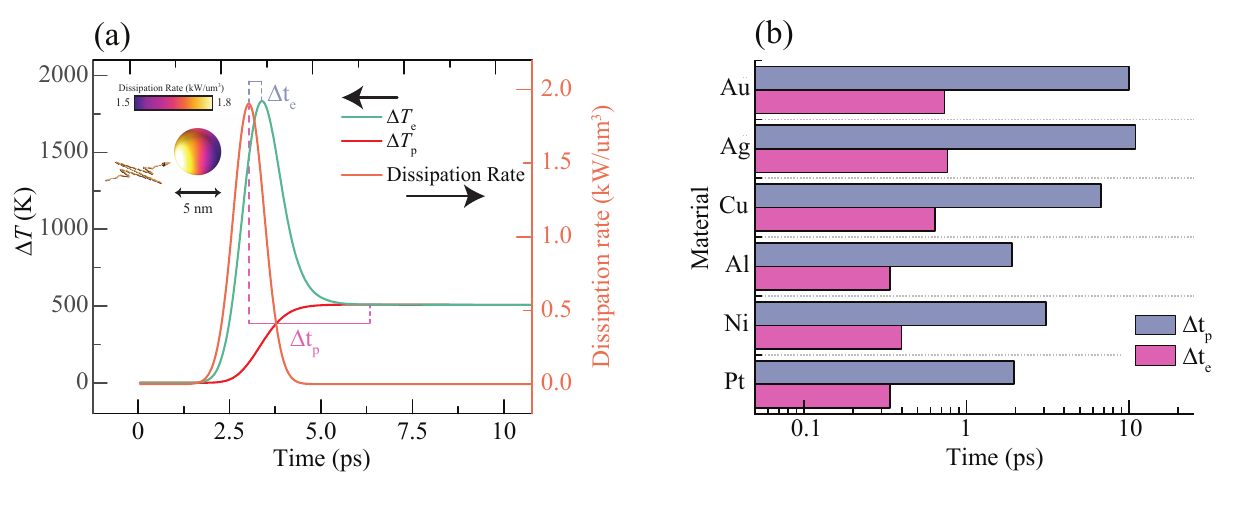}
    \caption{
        (a) Temporal evolution of the electron (\( \Delta T_e \)) and phonon (\( \Delta T_p \)) temperature increases and the dissipation rate in a 5 nm platinum sphere under femtosecond laser excitation (wavelength = 600 nm, pump pulse duration = 1 ps). The dissipation rate, with a maximum value of \( 1.8 \, \mathrm{kW/\mu m^3} \), is plotted on the right axis, while the temperature increases are shown on the left axis. Key relaxation times (\( \Delta t_e \) and \( \Delta t_p \)) are highlighted with dashed lines. The inset schematically depicts the excitation by an electromagnetic wave and the dissipation rate within the 5 nm Pt sphere, with dissipation values ranging from \( 1.5 \, \mathrm{kW/\mu m^3} \) to \( 1.8 \, \mathrm{kW/\mu m^3} \). 
        (b) Comparison of relaxation times (\( \Delta t_e \) and \( \Delta t_p \)) across various metallic materials (Au, Ag, Cu, Al, Ni, Pt). Bars represent these timescales, showcasing material-dependent differences in ultrafast thermal dynamics. The logarithmic time scale emphasizes the broad variation in response times, highlighting Pt's rapid electron-lattice coupling.
    }
    \label{fig2}
\end{figure}

As shown in Fig.~\ref{fig2}, the orange curve represents the temporal evolution of the dissipation rate under a femtosecond laser pulse (600 nm wavelength and 1 ps duration). The dissipation rate is fixed with a maximum value of $1.8 \, \mathrm{kW/\mu m^3}$. In panel (a), we plot $T_e$ and $T_p$ for platinum (Pt), a material with strong electron-phonon coupling. 
We define the characteristic relaxation times $\Delta t_e$ and $\Delta t_p$ as the intervals between the dissipation peak and the respective temperature peaks. 
The electron temperature reaches its maximum rapidly, whereas the phonon temperature responds more slowly, approaching its peak value after a delay.

Figure~\ref{fig2}(b) compares \(\Delta t_e\) and \(\Delta t_p\) across different metals, revealing a strong material dependence. Metals such as Pt, Ni, Al, and Cu exhibit relatively small \(\Delta t_e\), indicating faster electronic response times due to their higher electron-phonon coupling coefficients. 
In contrast, Au and Ag display longer \(\Delta t_e\), corresponding to slower relaxation. 
These trends suggest that the electron relaxation rate strongly influences the occurrence and magnitude of transient temperature inversion ($T_p > T_e$), a phenomenon that we will examine in detail using our 3D model.

\begin{figure}[htbp]
    \centering
    \includegraphics[width=1\textwidth]{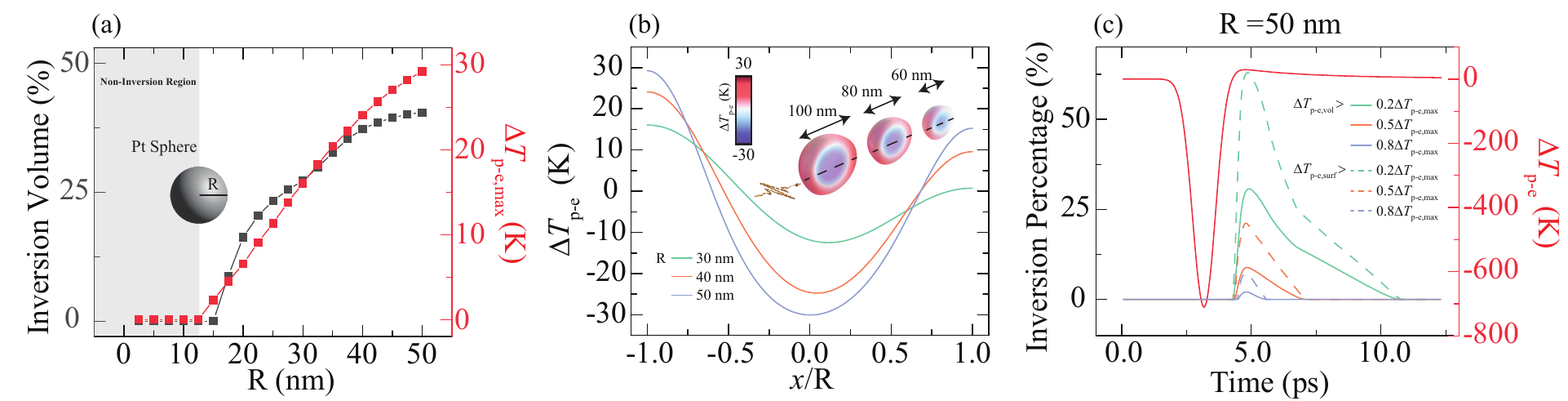}
    \caption{
        (a) Dependence of the inversion volume percentage (black squares, left axis) and maximum temperature difference (\( \Delta T_{p-e,\text{max}} \), red squares, right axis) on the radius (\( R \)) of a platinum sphere under femtosecond laser excitation (wavelength = 600 nm, pump pulse duration = 1 ps). The inversion volume percentage in this panel represents the proportion of the sphere's volume where \( \Delta T_{p-e} \) exceeds 10\% of \( \Delta T_{p-e,\text{max}} \). The critical radius (\( R_c = 12.5 \, \mathrm{nm} \)) separates the non-inversion region (gray-shaded area) from the inversion region. The inset illustrates the geometry of the Pt sphere. 
        (b) Spatial distribution of the electron-phonon temperature difference (\( \Delta T_{p-e} \)) across the sphere for various radii (\( R = 30, 40, 50 \, \mathrm{nm} \)) at the time of maximum inversion. The curves in this panel are taken along the dashed line suggested in the inset, plotting \( \Delta T_{p-e} \) versus the normalized radial position (\( x/R \)). The inset also shows the color map of \( \Delta T_{p-e} \) across the sphere, with the orange curve indicating the incident laser direction. The results indicate that the surface exhibits larger temperature differences than the interior, suggesting that inversion is more likely to occur at the surface.
        (c) Temporal evolution of the inversion volume percentage (solid lines, left axis) and surface inversion percentage (dashed lines, left axis) for a sphere with \( R = 50 \, \mathrm{nm} \). The solid lines represent the percentage of the sphere's volume where \( \Delta T_{p-e} \) exceeds specific thresholds (\( 0.2, 0.5, \text{and } 0.8 \) of \( \Delta T_{p-e,\text{max}} \)), while the dashed lines represent the corresponding thresholds for the surface. These results emphasize the surface-dominated nature of temperature inversion, where surface dissipation gradients enhance the likelihood and magnitude of inversion.
    }

    \label{fig3}
\end{figure}

For such a small nanostructure, the electronic temperature, $T_e$ is always larger than the phononic one, $T_p$ and no temperature inversion is observed ($\Delta T_{p-e}<0$). 

To understand the pivotal role of geometric size and dissipation gradients in dictating the temperature inversion behavior, we calculate the maximum value over time of $\Delta T_{p-e}$ reached in Pt nanospheres of different sizes. The results presented in Fig.~\ref{fig3} show the existence of a critical size value ($R_c = 12.5 \, \mathrm{nm}$), below which the dissipation rate pattern is homogeneous enough to effectively suppress any significant temperature inversion (Fig.~\ref{fig3} right axis). In contrast, for larger spheres ($R > R_c$), the dissipation rate becomes increasingly inhomogeneous, particularly near the surface, leading to pronounced temperature inversion. Along with the magnitude of $\Delta T_{p-e}$, larger particles also exhibit larger fractions of volume undergoing temperature inversion (Fig.~\ref{fig3} left axis). This transition is visually supported by the dissipation rate distributions in Supplementary Information (Fig.~S1), where larger particles ($R = 45 \, \mathrm{nm}$) exhibit substantial dissipation gradients, while smaller particles ($R = 5 \, \mathrm{nm}$) show more uniform dissipation, damping the mechanism underlying the emergence of temperature inversion, studied in detail in the following.

As shown in Fig.~\ref{fig3}(b), temperature inversion is predominantly observed at the surface of the sphere, where dissipation gradients are most pronounced. The red regions in the inset of Fig.~\ref{fig3}(b) correspond to higher temperature inversion values, indicating a clear surface-dominated behavior. Additionally, Fig.~\ref{fig3}(c) demonstrates that the surface inversion percentage is consistently higher than the volume inversion percentage across all thresholds, reinforcing the role of surface regions as the primary site of temperature inversion. This observation can be relevant for surface-related phenomena, such as adsorption/desorption mechanisms during photocatalysis which may be affected by the direction of energy transfer between electrons and phonons. The similarity in end times for both surface and volume inversion in Fig.~\ref{fig3}(c) further highlights that radius-dependent dissipation gradients, rather than temporal differences, drive these dynamics.

It should be noted that the temperature inversion phenomenon is not a resonant effect. At $\lambda = 600$~nm, neither the 5 nm nor the 45 nm radius particles are at resonance (Fig.~S2a).  While the absorption efficiency ($\eta_{\mathrm{abs}}$) at 600 nm increases for larger radii (Fig.~S2a), its absolute value remains small, and the irradiation fluence required to achieve the presented temperature inversions can be quite large. It would be, therefore, relevant to understand how temperature inversion can be achieved under milder conditions. One possibility is to introduce designs featuring geometrical asymmetries that explicitly display inhomogeneous dissipation patterns, as shown in the following.

\subsection{3D Bowtie nanostructure}
We now examine a more complex 3D bowtie geometry to explore how spatial inhomogeneities (both geometrical and optical-related), material properties, and excitation conditions jointly influence ultrafast temperature inversion. This approach allows us to investigate not only temporal trends but also how inversion emerges, evolves, and fades across the nanostructure’s spatial extent.

\begin{figure}
    \centering
    \includegraphics[width=1\textwidth]{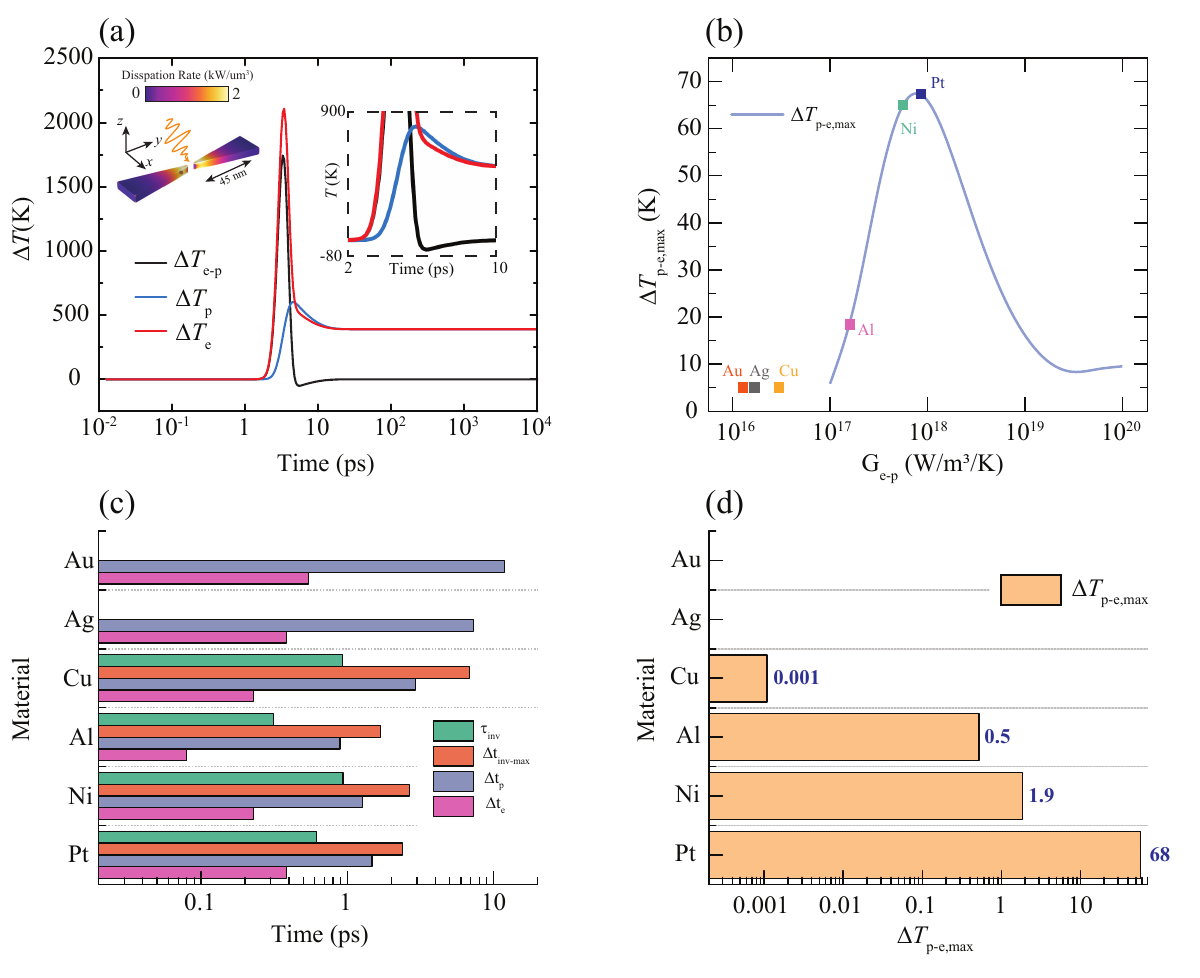}
    \caption{
    (a) Temporal evolution of electron (\( T_e \)) and lattice (\( T_p \)) temperatures for an extended bowtie nanostructure illuminated by a femtosecond laser pulse (600 nm wavelength, 75 \(\mu\mathrm{J}/\mathrm{cm}^2\) fluence, 1 ps duration). A temperature inversion occurs near the bow-tie tips, as highlighted in the inset, which zooms into the critical time range.
    (b) Dependence of the maximum temperature inversion (\( \Delta T_{p,\text{max}} \)) on the electron-phonon coupling coefficient (\( G_{\text{ep}} \)) for various materials. Platinum (Pt) exhibits the most pronounced inversion, followed by Nickel (Ni) and Aluminum (Al).
    (c) Comparative analysis of material-dependent thermal responses, showing inversion duration, time to maximum inversion, and time-to-peak values of \( T_{p-e} \), \( T_e \), and hot electron density (\( \rho_{\mathrm{HE}} \)).
    (d) Bar chart of the maximum temperature inversion (\( \Delta T_{p,\text{max}} \)) for different materials. Pt exhibits the highest inversion, while Copper (Cu) shows negligible values.
    }
    \label{fig4}
\end{figure}

As shown in Figure~\ref{fig4}(a), the brown point close to the tip in a 3D Pt bowtie (45 nm long, 20 nm wide and 5 nm thick) under illumination with a femtosecond pulse (600 nm wavelength, $75 \, \mu\mathrm{J}/\mathrm{cm}^2$ fluence, and 1 ps duration) exhibits a clear temperature inversion  $(T_p > T_e)$. 
Here, we quantify key timescales defined earlier: \(\Delta t_e\) (time from dissipation peak to electron temperature peak), \(\tau_{\mathrm{inv,max}}\) (time from dissipation peak to maximum inversion), and \(\tau_{\mathrm{inv}}\) (inversion duration from onset \(T_p-T_e=0\) to its peak). 
Note that \(\tau_{\mathrm{inv}}\) characterizes the duration of the inversion, providing a direct measure of its stability.

Figure~\ref{fig4}(b) links the maximum inversion \(\Delta T_{p,\text{max}}\) to the electron-phonon coupling coefficient \(G_{e-p}\), while keeping the other material parameters as in Pt. 
We find that the intrinsic properties of Pt yield the strongest inversion, whereas other metals show smaller values. 
Figure~\ref{fig4}(d) summarizes this material ranking, and Figure~\ref{fig4}(c) examines temporal metrics (\(\tau_{\mathrm{inv}}, \tau_{\mathrm{inv,max}}\)) for each metal. 
As observed for the spherical case, strong electron-phonon coupling and rapid electron relaxation (Pt, Ni, Al, Cu) favor inversion, while weaker coupling leads to negligible inversion (Au, Ag). 
This confirms the material-dependent patterns observed previously and recasts them in a more complex geometry, importantly showing how sizable temperature inversions (tens of degrees) can be achieved with relatively low fluences (e.g., $75 \, \mu\mathrm{J}/\mathrm{cm}^2$)

\begin{figure}
    \centering
    \includegraphics[width=0.75\textwidth]{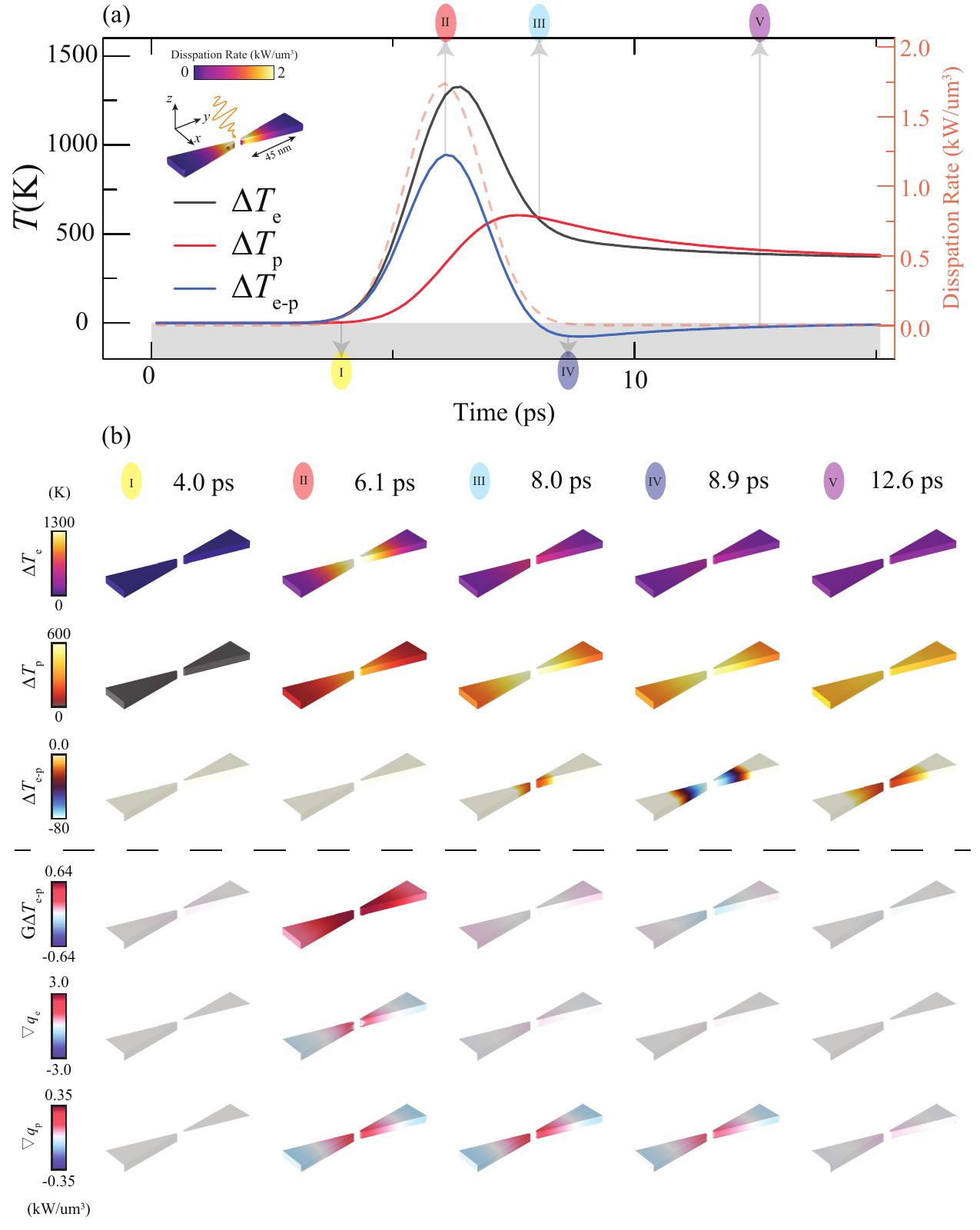}
    \caption{
    Thermal dynamics of the bowtie nanostructure under femtosecond pulsed laser illumination. 
    (a) Temporal evolution of the electronic temperature (\( T_e \)), lattice temperature (\( T_p \)), and the electron-lattice temperature difference (\( \Delta T_{\text{e-p}} = T_e - T_p \)). The orange dashed curve shows the dissipation rate of the illumination at a specific location in the bowtie nanostructure (second y-axis). Key time points (\( I \)–\( V \)) are indicated for further spatial analysis. 
    (b) Spatial distributions of \( T_e \), \( T_p \), \( \Delta T_{\text{e-p}} \), the electron-phonon coupling term (\( G_{\text{e-p}} \Delta T_{\text{e-p}} \)), and heat flux divergences (\( \nabla \cdot q_e \) and \( \nabla \cdot q_p \)) at selected times (4.0\,ps, 6.1\,ps, 8.0\,ps, 8.9\,ps, and 12.6\,ps). These distributions illustrate the dominant role of electron-phonon coupling (\( G_{\text{e-p}} \)) and electron heat diffusion in the thermal response, with minimal contribution from lattice heat diffusion (\( \nabla \cdot q_p \)).
    }
    \label{fig5}
\end{figure}

To clarify what fundamental mechanisms lead to temperature inversions, Figure~\ref{fig5} provides a spatio-temporal grid showing when and where inversion emerges. 
Initially ($\sim$ 4 ps), electron and lattice temperatures are closely matched. By $\sim$6 ps, \(\Delta T_{e-p}\) is maximized, and both electron-phonon coupling and \(\nabla \cdot q_e\) peak—electrons are significantly transferring energy to the lattice and through the colder electronic surroundings, respectively. 
Between 6 and 8 ps, although the lattice temperature contour remains almost unchanged, the electron-driven energy redistribution tapers off as \(\nabla \cdot q_e\) decreases. 
Around 8 ps, the system crosses into the inversion regime (\(T_p-T_e>0\)).

By $\sim$ 9 ps, spatial maps show clearly demarcated inversion zones. 
At this juncture, the electron-phonon coupling term changes sign, indicating a reversal of energy flow (the lattice returning energy to the electrons) and a gradual diminishment of the inversion magnitude. 
These spatial results confirm that inversion is not uniform: it arises in localized regions and then recedes as coupling conditions evolve and heat flux gradients weaken. 
Thus, the spatial distributions reinforce the temporal metrics, demonstrating that both time and space-related effects govern the inversion’s intensity and lifetime.

\begin{figure}
    \centering
    \includegraphics[width=1\textwidth]{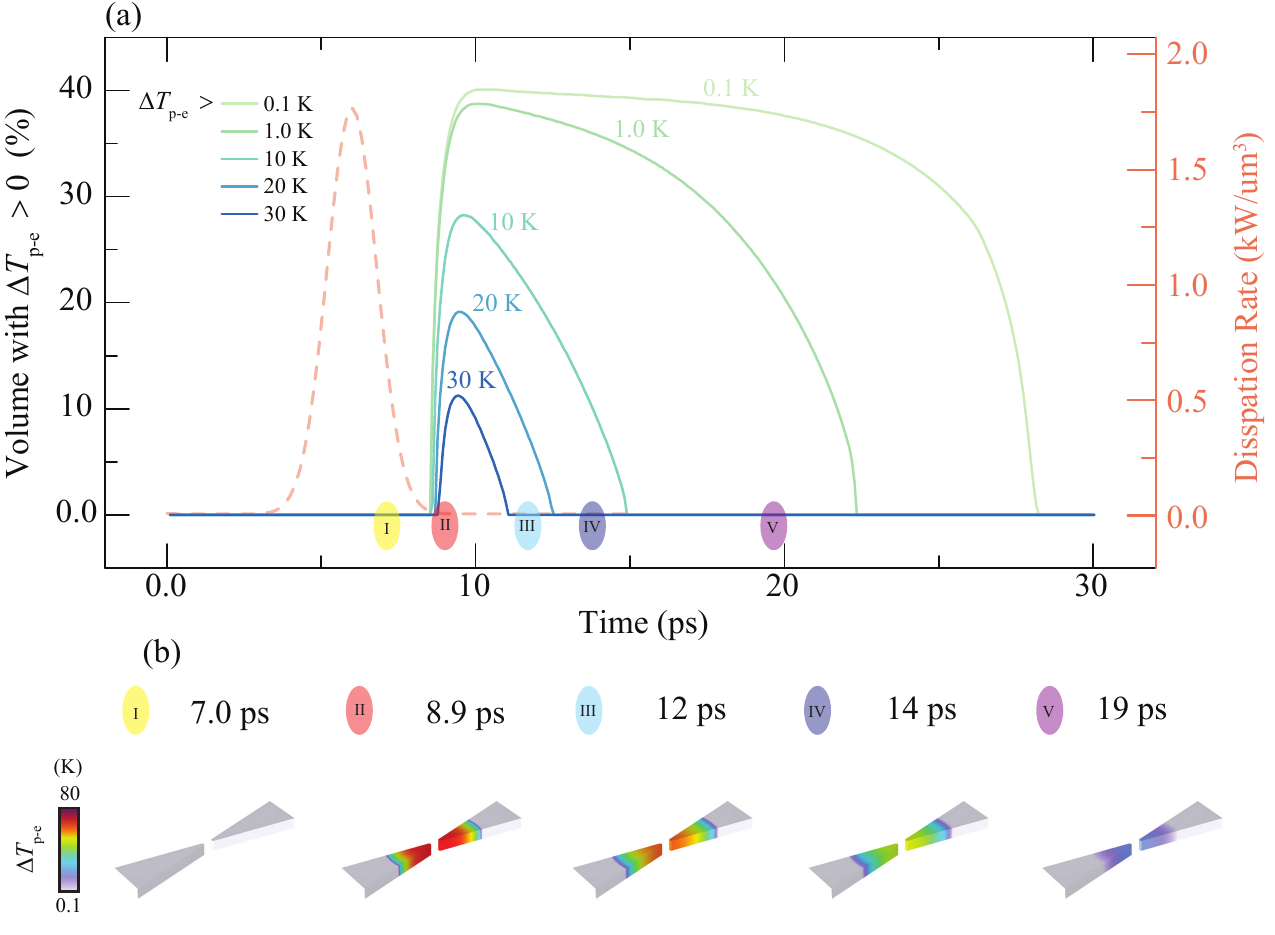}
    \caption{
    (a) Temporal evolution of the percentage of the bowtie nanostructure exhibiting a temperature inversion (\( \Delta T_{e-p} = T_e - T_p \)) under pulsed laser illumination. The curves represent the proportion of the structure exceeding different inversion thresholds (\( \Delta T_{e-p} > 0.1, 1, 10, 20, 30 \,\mathrm{K} \)) over time. The orange dashed curve shows the dissipation rate of the illumination at a specific location in the bowtie nanostructure. Simulation parameters: wavelength = 600\,nm, pulse fluence = 75\,\(\mu\mathrm{J}/\mathrm{cm}^2\), and pulse duration = 2\,ps. Key time points (I--V) are labeled for spatial analysis. 
    (b) Spatial distributions of the temperature difference (\( \Delta T_{e-p} \)) at selected times (7.0\,ps, 8.9\,ps, 12\,ps, 14\,ps, and 19\,ps), corresponding to the marked points in (a). The color bar indicates the magnitude of \( \Delta T_{e-p} \), highlighting the temporal evolution of thermal gradients across the bowtie structure.
    }
    \label{fig6}
\end{figure}

To further quantify the evolving spatial extent and intensity of inversion, Figure~\ref{fig6}(a) tracks the fraction of the nanostructure volume surpassing various inversion thresholds. 
Initially, low-level inversion (\(>0.1\,\mathrm{K}\)) is widespread, suggesting the phenomenon begins as a moderate imbalance.
Over time, smaller regions sustain higher inversions (e.g., $>20\,\mathrm{K}$), culminating in transient hotspots of intense inversion that eventually subside. 
The snapshots in Figure~\ref{fig6}(b) highlight this progression: from weak, broadly distributed inversion (I) to well-defined, high-inversion locales at 8.9 ps (II), then a retreat back toward equilibrium (III–V). 
These temporal and spatial data together illustrate that inversion intensity peaks in localized spots and does not uniformly affect the entire nanostructure at once, mirroring the material-dependent patterns and coupling-driven flows discussed earlier.\\

\begin{figure}
    \centering
    \includegraphics[width=1\textwidth]{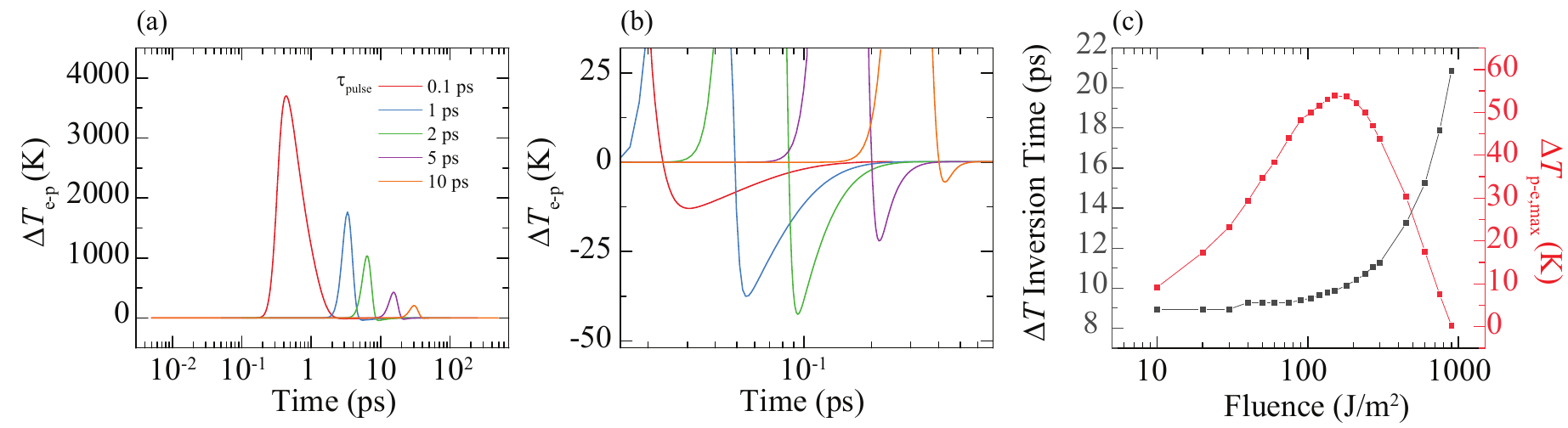}
    \caption{
    (a) Temporal evolution of the electron–phonon temperature difference 
    ($\Delta T_{\mathrm{e-p}} = T_e - T_p$) for various pump durations 
    ($\tau_{pulse} = 0.1\,\mathrm{ps}, 1\,\mathrm{ps}, 2\,\mathrm{ps}, 5\,\mathrm{ps}, 10\,\mathrm{ps}$). 
    The curves display peak structures that depend strongly on the pump duration, highlighting the transient dynamics of the system.  
    (b) A closer view of the temperature inversion reveals that the magnitude increases 
    with $\tau_{pulse}$, reaching a maximum near $2\,\mathrm{ps}$ before decreasing at longer 
    durations (5 and 10 ps), indicating an optimal pump duration for maximizing inversion. 
    (c) Dependence of inversion time (black squares) and maximum temperature difference 
    (red squares) on fluence (600 nm wavelength, 2 ps duration). The inversion time increases monotonically with fluence, 
    while the maximum temperature difference exhibits a nonmonotonic behavior, peaking at 
    intermediate fluences.}
    \label{fig7}
\end{figure}

Beyond material factors and spatial patterns, the temporal profile of photon energy absorption also plays a role.
Figure~\ref{fig7} shows how varying the pump duration \(\tau_{pulse}\) modulates \(\Delta T_{e-p}\).
Extremely short pulses (0.1 ps) deposit energy abruptly, creating a large but fleeting imbalance.
Longer pulses (5–10 ps) never reach such a steep non-equilibrium state, resulting in milder inversions.
Remarkably, an intermediate duration (\(\tau_{pulse} = 2\,\mathrm{ps}\)) produces the most robust and long-lived inversion. 
This aligns with the previous observations: the conditions for strong inversion require a moderate balance in how and when energy is supplied to the electron-phonon system.

Finally, Figure~\ref{fig7} highlights fluence as another tuning parameter. 
As fluence increases, the inversion peak shifts to later times, indicating that more energy input delays the maximum imbalance. 
However, \(\Delta T_{p-e,\text{max}}\) exhibits a nonmonotonic trend, peaking at an intermediate fluence before declining. 
This complexity arises from temperature-dependent electron heat capacity and thermal conductivity. 
If these parameters were constant and independent of temperature (see SI for a fixed-parameter analysis), no peak would occur. 
Instead, the temperature dependence enhances the inversion at a certain energy level, then suppresses it as the system’s relaxation pathways become more effective.

In summary, we see that material properties dictate the potential for inversion, spatial distributions confirm that inversion is both localized and transient, and adjusting pump duration or fluence can fine-tune the strength and timing of the phenomenon. 
Moreover, the nonmonotonic behavior at certain fluences underscores the importance of accurate temperature-dependent material parameters. 

\subsection{Comparison of 0D and 3D Models}

Having explored how geometry, material parameters, and excitation conditions impact the temperature inversion in the 3D bowtie nanostructure, to capture the essential mechanisms underlying temperature inversion effects, we now compare these detailed simulations with a reduced 0D approach.  
As introduced in the Methods section, the 0D model captures essential electron-phonon energy exchange physics by treating the system as spatially uniform and incorporating an effective diffusive loss term \((k_{e,p} / (b H)^2)\), where $b$ is a fitting parameter and $H$ is the bowtie's height.

By selecting $b = 0.45$, we determine that $b H$ serves as the effective characteristic length, bridging the 0D and 3D descriptions.  
As detailed in the SI, this fitting parameter is optimized by analyzing the 3D simulation results and identifying the value that minimizes discrepancies between the 0D and 3D outcomes.  
The resulting effective length scale, $b H$, ensures that the 0D model faithfully reproduces the electron-lattice energy exchange and relaxation behavior observed in the full 3D simulations.

While the 0D framework lacks spatial resolution, it significantly reduces computational complexity, enabling rapid parameter scans and initial explorations of how material properties, electron-phonon coupling coefficients and pulse conditions influence the inversion magnitude.  
The use of $b H$ as the effective length provides a systematic and reliable approach for mapping the detailed 3D results onto the simplified 0D model.

\begin{figure}
    \centering
    \includegraphics[width=1\textwidth]{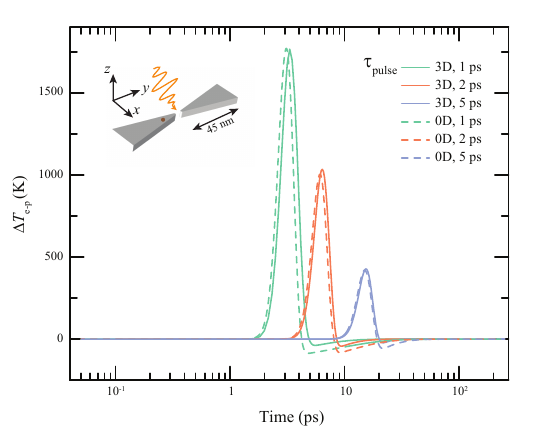}
    \caption{
    Temporal evolution of the electron-phonon temperature difference (\(\Delta T_{\mathrm{e-p}} = T_e - T_p\)) computed at the indicated location (brown dot in the inset), for three different pump durations ($\tau_{pulse} = 1~\mathrm{ps}, 2~\mathrm{ps}, 5~\mathrm{ps}$; green, orange, blue). Solid lines represent results from three-dimensional (3D) simulations, while dashed lines show predictions from the simplified zero-dimensional (0D) model. Despite minor deviations in peak positions and amplitudes, the 0D model closely reproduces the key features and magnitudes observed in the more detailed 3D calculations, effectively capturing the essential temperature inversion dynamics over the relevant timescales.}
    \label{fig8}
\end{figure}

As shown in Figure~\ref{fig8}, the 0D analytical and 3D numerical models yield similar inversion profiles, with only minor discrepancies in peak timing and amplitude. 
This close agreement suggests that the fundamental physics—namely, the interplay between electron-phonon coupling, pulse-driven non-equilibrium electron heating, and subsequent relaxation—are not highly sensitive to spatial complexity once an appropriate length scale is chosen. 
The 0D model thus serves as a valuable proxy, capturing the main temporal characteristics of the inversion without the computational expense of fully resolved 3D simulations.

\begin{figure}[htbp]
    \centering
    \includegraphics[width=1\textwidth]{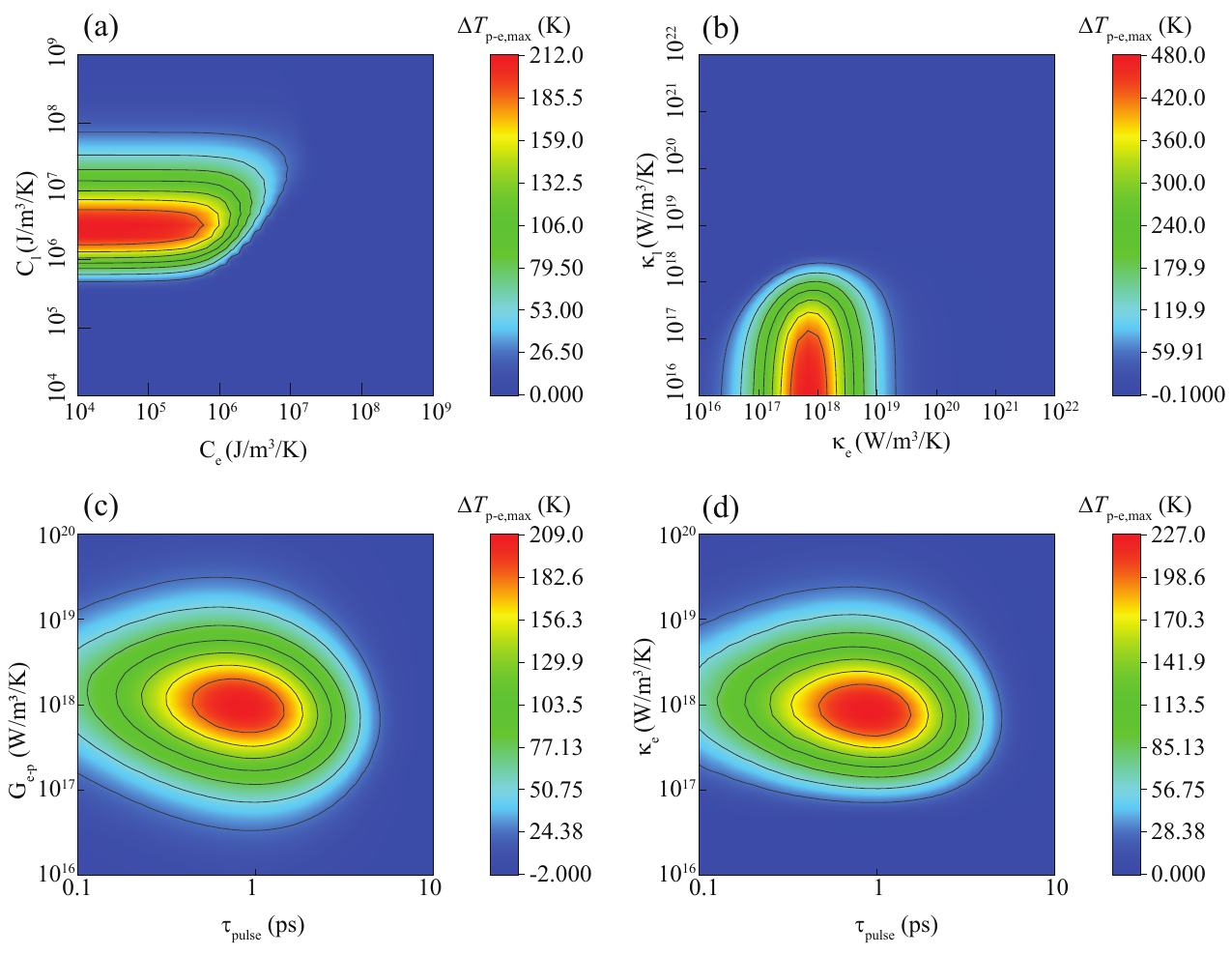} 
    \caption{Contour plots of the maximum phonon–electron temperature inversion, \(\Delta T_{p-e,\text{max}} = \max(T_p - T_e)\), as a function of various material and pulse parameters in the 0D electron–phonon coupling model. (a) Variation with electronic heat capacity (\(C_e\)) and lattice heat capacity (\(C_p\)) demonstrates that moderate values of both parameters promote pronounced inversions, highlighting the balance between energy storage and dissipation. (b) Variation with electronic (\(\kappa_e\)) and lattice (\(\kappa_p\)) heat diffusions reveals that intermediate values optimize the inversion by maintaining an equilibrium between heat retention and dissipation. (c) Variation with electron–phonon coupling strength (\(G_{\text{e-p}}\)) and pulse duration (\(\tau_{\text{pulse}}\)) shows that stronger coupling, combined with shorter pulses, enhances inversion by increasing energy deposition rates. (d) Variation with \(\kappa_e\) and \(\tau_{\text{pulse}}\) further underscores the critical role of thermal transport and pulse characteristics in determining the magnitude of \(\Delta T_{p-e,\text{max}}\). These results provide insights into tailoring ultrafast material responses for applications in nanoscale heat management and energy transfer. Further methodological details, including governing equations and parameter descriptions, are provided in Section Zero-Dimensional Analytical Framework.}
    \label{fig:0D_coupling}
\end{figure}

Incorporating the results shown in Fig.~\ref{fig:0D_coupling}, we exploit the efficiency of the 0D approach to perform extensive parameter sweeps over material properties (e.g., \(C_e\), \(\kappa_e\), \(G_{e-p}\)) and pulse parameters (\(\tau_{\text{pulse}}\)) to gain insights over the mechanisms governing the temperature inversion process. For these calculations, Pt material properties and a pulse duration of 1 ps were utilized for non-varying parameters. The results confirm and extend the trends observed in both the 3D simulations and the analytical model, revealing optimal parameter regimes for achieving pronounced temperature inversions.

As shown in Fig.~\ref{fig:0D_coupling}(a), a specific interval of lattice heat capacities values (\(C_p\)) provide the highest temperature inversion, provided that the electronic heat capacity (\(C_e\)) is small enough. A too large or too small \(C_p\) would lead to too small or too early $\Delta T_p$ increase while a too large $C_e$ would limit and delay $\Delta T_e$ rise.  This trend highlights the importance of balancing energy storage and dissipation, where excessively high or low values lead to suboptimal energy transfer dynamics. Similarly, Fig.~\ref{fig:0D_coupling}(b) illustrates the effect of electronic (\(\kappa_e\)) and lattice (\(\kappa_p\)) heat diffusion, revealing that intermediate values optimize inversion by maintaining an equilibrium between heat retention and dissipation. If $\kappa_e$ is too large, $T_e$ spatially redistributes before increasing enough locally. If $\kappa_e$ is too small, spatial electronic heat diffusion is damped and no inversion can occur, similarly to what happens in the small particle case. Regarding lattice thermal conductivity, as long as $\kappa_p$ is small enough, $T_p$ can locally increase and lead to temperature inversion as soon as $T_e$ has locally dropped, thanks to electronic heat diffusion.

The importance of a balance between heat transport dynamics and temperature increase is more deeply understood in  Fig.~\ref{fig:0D_coupling}(c,d) where electron–phonon coupling strength (\(G_{e-p}\)) and electronic heat diffusion (\(\kappa_e\)) are varied against pulse duration ($\tau_{\text{pulse}}$). If $\tau_{\text{pulse}}$ is too large, electrons are heated for longer times preventing $T_p$ from overtaking $T_e$, and the temperature inversion is lost, as expected. Interestingly while temperature inversion persists down to 100 fs irradiation times, the maximum $\Delta T_{p-e}$ is achieved for ps-long pulses, further underscoring the importance of thermal transport and pulse characteristics in determining the magnitude of \(\Delta T_{p-e,\text{max}}\).

Another key insight comes from comparing Fig.~\ref{fig:0D_coupling}(c) and Fig.~\ref{fig:0D_coupling}(d) more closely. Given a pulse duration, the optimal values for \(G_{e-p}\) and \(\kappa_e\) to maximize \(\Delta T_{p-e}\) are very close to each other. This is not a coincidence as temperature inversion relies on the balance between power transfer across electron and phonons, in parallel to electronic heat diffusion. This mechanism allows the lattice temperature to increase at the same rate as the electronic temperature locally decreases.

Intuition for this dynamics can be captured utilizing an even simpler 0D model where we assume (i) similar electronic and lattice heat capacities, $C_e\approx C_p=C$ (e.g., at the beginning of the optimal interval of Fig.~\ref{fig:0D_coupling}(a)), (ii) a relatively small lattice heat diffusion $\kappa_p<<\kappa_e$ (consistent with $k_e\approx k_p\frac{T_e}{T_p}$ at high electronic temperatures) and (iii) a short pulse temporally approximated by a Dirac delta $\delta (t)$.

Under these assumptions, the 0D model simplifies to:

\begin{equation}
\begin{cases}
C\dot{T_e}(t)=-G_{e-p}\Delta T_{e-p}(t)-\kappa_e T_e(t)+Q_0\tau \sqrt{2\pi}\delta(t-t_0)\\
C\dot{T_p}(t)=G_{e-p}\Delta T_{e-p}(t)
\end{cases}
\end{equation}

where $\tau$ is the equivalent pulse duration for a non $\delta(t)$-like pulse with an associated power dissipation density of $Q_0e^{-\frac{(t-t_0)^2}{4\tau^2}}$.

 This system has a closed-form solution for the temperature inversion $\Delta T_{p-e}(t)$ which, although of relatively complex form (see SI for details), can be maximized at its peak if $G_{e-p}=\kappa_e/2$. This analytical result is consistent with the above discussion and Fig.~\ref{fig:0D_coupling}(c,d) and explains why only some materials allow sizable temperature inversions for the investigated nanostructures. Further studies should focus on the tunability of a particle geometry to achieve electronic heat diffusion characteristic lengths yielding $G_{e-p}\approx \kappa_e$. This strategy may add another knob to manipulate electron-phonon energy exchange dynamics.

 While 3D simulations are crucial for capturing spatial distributions and refining quantitative accuracy, the 0D approximations offer a powerful tool for guiding initial searches in parameter space. 
 Together, they establish a complementary framework for understanding, predicting, and ultimately controlling ultrafast temperature inversion phenomena in complex plasmonic nanostructures.

\section{Conclusions}

In this study, we have comprehensively investigated the ultrafast thermal response of metallic nanostructures illuminated by femtosecond laser pulses, bridging the gap between zero-dimensional (0D) analytical models and fully three-dimensional (3D) finite-element simulations. 
Our findings reveal that the emergence, strength, and duration of transient temperature inversion, where the lattice temperature temporarily surpasses the electron temperature, depend sensitively on material properties (e.g., electron-phonon coupling and thermal conductivities), geometrical configuration, and excitation parameters (fluence and pulse duration).

Among the materials considered, platinum (Pt) consistently exhibits the most pronounced and sustained inversion, whereas other metals show weaker or negligible effects. 
Additionally, we identify an optimal fluence range and intermediate pulse durations that maximize the inversion, emphasizing the nontrivial interplay between energy input, relaxation mechanisms, and temperature-dependent material parameters. 
Spatially resolved 3D simulations confirm that inversion occurs in localized hotspots and is not uniform across the nanostructure, while the simpler 0D framework, when supplied with a suitable characteristic length, captures the essential temporal behavior and guides rapid parametric explorations.

These insights offer a unifying perspective on how to manipulate ultrafast temperature inversion phenomena through material selection, nanoscale design, and tailored excitation conditions. 
By integrating both 0D and 3D models, the parameter space can be efficiently navigated, identifying optimal regimes, and ultimately informing the engineering of nanostructures for hot-carrier generation, photothermal processes, and nanoscale heat management. 
Moreover, experimental investigations could be envisaged to track the temperature inversion by leveraging time-resolved nonlinear optical techniques. For instance, ultrafast pump-probe spectroscopy has proven to be a powerful tool to unveil the dynamics of plasmonic nanostructures \cite{BDC2011,H2011,BSN2017}, and also to distinguish between electronic and phononic contributions to the transient optical signal \cite{CCP2014,HHH2000,RDM2021}.

This improved understanding of the ultrafast electron-phonon coupling dynamics lays the groundwork for more predictive design strategies in applications ranging from energy harvesting and photothermal therapies to advanced pulsed-regime plasmonic device technologies.

\section*{Acknowledgments}

Q.Y. acknowledges support from the Robert A.Welch Foundation under grant C-1222.

S.K.S. acknowledges support from the Army Research Laboratory under Cooperative Agreement Number W911NF24-2-0060.

A.S. acknowledges the European Union’s Horizon Europe research and innovation programme under the Marie Sk\l{}odowska-Curie Action PATHWAYS HORIZON-MSCA-2023-PF-GF grant agreement no.~101153856.

A.A. acknowledges support from US Army grant W911NF2420105

\bibliographystyle{unsrt}
\bibliography{refs}

\clearpage
\section*{Supporting Information}

\setcounter{figure}{0}
\renewcommand{\thefigure}{S\arabic{figure}}
\setcounter{table}{0}
\renewcommand{\thetable}{S\arabic{table}}

\begin{figure}[hbtp]
    \centering
    \includegraphics[width=1\textwidth]{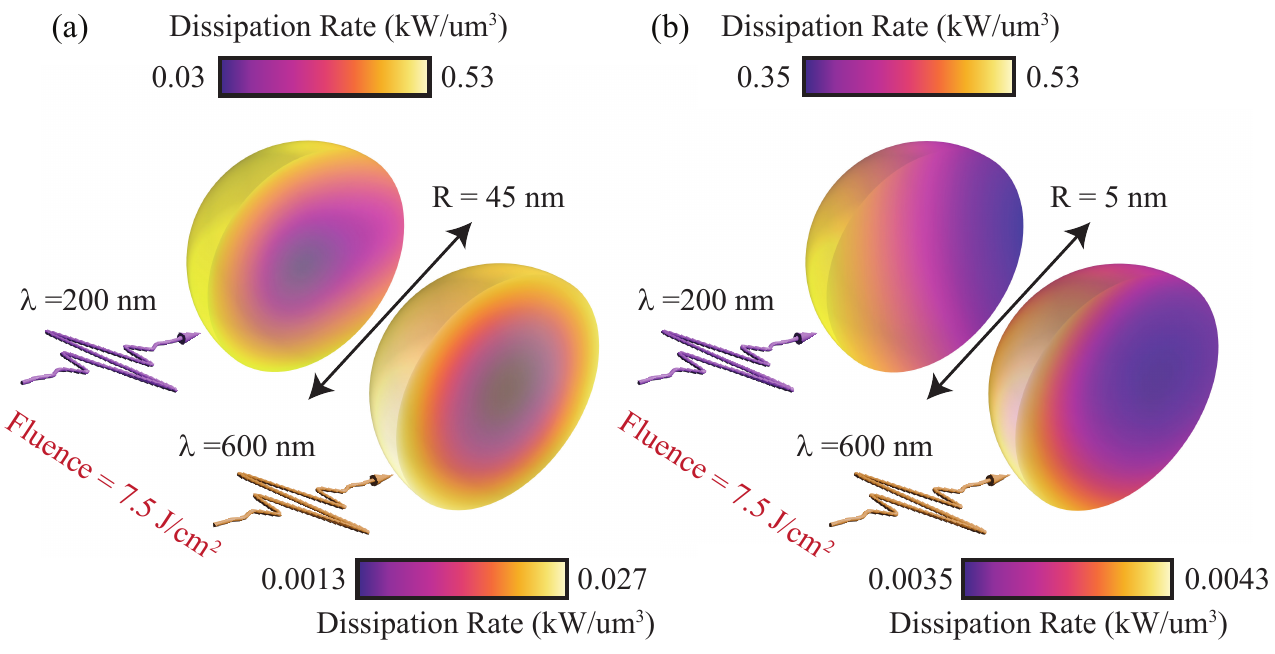}
    \caption{ Dissipation rate (\(\text{kW}/\mu\text{m}^3\)) distributions for spherical nanoparticles under illumination at two wavelengths (\(\lambda = 200~\text{nm}\) and \(\lambda = 600~\text{nm}\)) and two radii (\(R = 45~\text{nm}\) and \(R = 5~\text{nm}\)) at a fluence of \(7.5~\text{J/cm}^2\). (a) For \(R = 45~\text{nm}\), the dissipation rate distribution is highly inhomogeneous, ranging from \(0.03\) to \(0.53~\text{kW}/\mu\text{m}^3\) at \(\lambda = 200~\text{nm}\), and from \(0.0013\) to \(0.027~\text{kW}/\mu\text{m}^3\) at \(\lambda = 600~\text{nm}\). This significant spatial inhomogeneity contributes to the temperature inversion phenomena observed in larger particles, as discussed in the main text. (b) For \(R = 5~\text{nm}\), the dissipation rate distribution is more homogeneous, with closer minimum and maximum values (\(0.35\) to \(0.53~\text{kW}/\mu\text{m}^3\) at \(\lambda = 200~\text{nm}\), and \(0.0035\) to \(0.0043~\text{kW}/\mu\text{m}^3\) at \(\lambda = 600~\text{nm}\)), reducing the likelihood of temperature inversion. These results underscore the critical role of particle size and illumination wavelength in determining energy dissipation patterns and their connection to thermal dynamics. Further details on the computational methodology are provided in Section Zero-Dimensional Analytical Framework of the main text. }
    \label{fig:dissipation_rate}
\end{figure}

\begin{figure}[hbtp]
    \centering
    \includegraphics[width=1\textwidth]{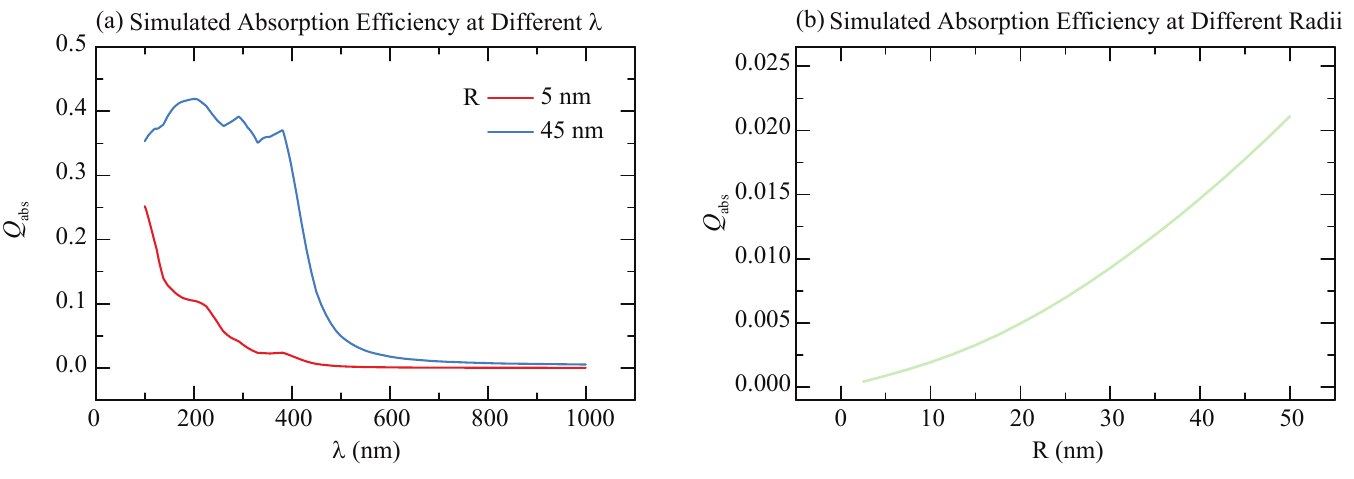}
    \caption{(a) Simulated absorption efficiency (\(Q_{\text{abs}}\)) as a function of wavelength (\(\lambda\)) for spherical nanoparticles with radii of \(R = 5~\text{nm}\) (red curve) and \(R = 45~\text{nm}\) (blue curve). The results demonstrate a pronounced inefficiency in absorption at \(\lambda = 600~\text{nm}\), particularly for larger radii, highlighting the challenges in achieving optimal energy conversion under specific illumination conditions. (b) Simulated absorption efficiency (\(Q_{\text{abs}}\)) as a function of particle radius (\(R\)) at a fixed wavelength of \(\lambda = 600~\text{nm}\), showing a monotonic increase in efficiency with increasing radius. These results provide a comprehensive view of how both wavelength and particle size influence absorption efficiency. The absorption efficiency was calculated using COMSOL simulations, with material properties provided in Tables S1 and S2 of the SI.}

\end{figure}

\begin{figure}[hbtp]
    \centering
    \includegraphics[width=1\textwidth]{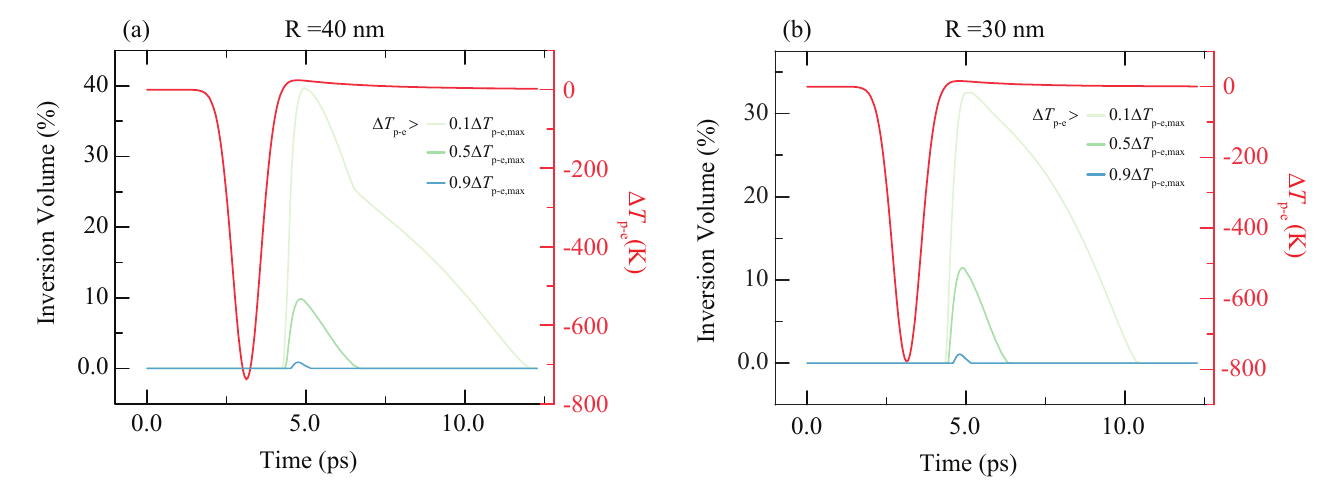} 
    \caption{Time evolution of the inversion volume percentage (left y-axis) and the phonon–electron temperature difference, \(T_{p-e}\) (right y-axis), for spherical nanoparticles of radii \(R = 40~\text{nm}\) (a) and \(R = 30~\text{nm}\) (b) in a 2D simulation. The inversion volume percentage represents the proportion of the nanoparticle's volume where the temperature difference, \(\Delta T_{p-e}\), exceeds specific thresholds (\(0.1\), \(0.5\), and \(0.9\) of the maximum \(\Delta T_{p-e}\)) over time. For \(R = 40~\text{nm}\), the inversion volume is larger and persists longer compared to \(R = 30~\text{nm}\), where it is significantly reduced and short-lived. The dynamics of \(T_{p-e}\) illustrate how particle size influences the temporal characteristics of energy transfer and thermal distribution. Further details on the 2D simulation methodology and parameter choices are provided in the main text.}
    \label{fig:2D-Percentage}
\end{figure}

\begin{figure}[hbtp]
    \centering
    \includegraphics[width=1\textwidth]{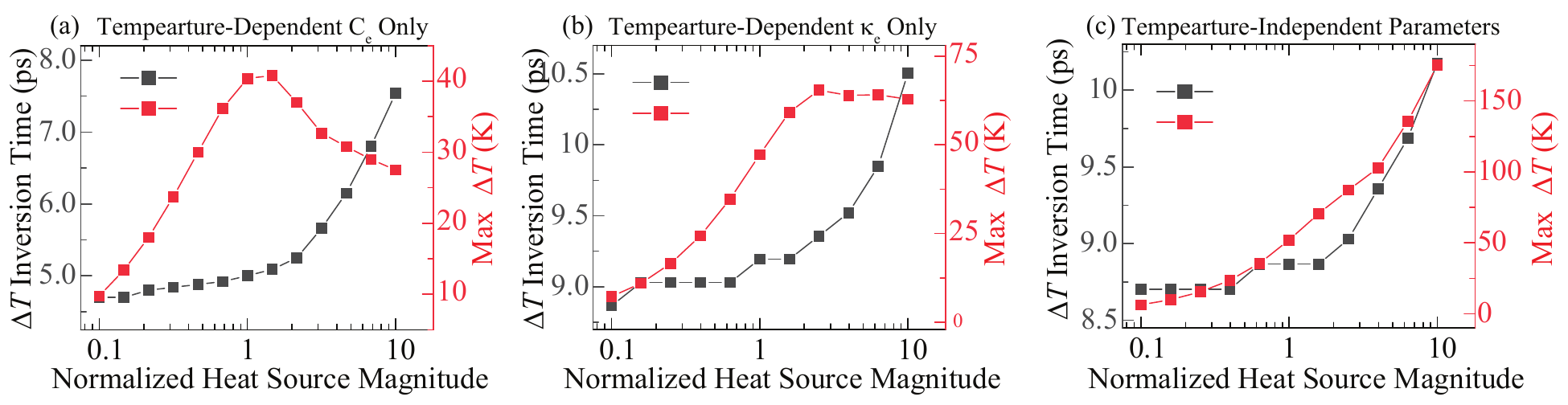}
    \caption{Influence of normalized heat source magnitude on $\Delta T$ inversion time and maximum $\Delta T$ for three different parameter configurations: (a) Temperature-dependent electron heat capacity ($C_e$) only, (b) temperature-dependent electron thermal conductivity ($\kappa_e$) only, and (c) temperature-independent parameters. The x-axis represents the normalized heat source magnitude, defined as a multiplicative factor applied to the heat source generated by a fluence of \(75~\mu\text{J/cm}^2\). The heat source is derived from a pump illumination at \(\lambda = 600~\text{nm}\) with a \(2~\text{ps}\) duration, forming a spatially and temporally dependent function determined by the pulse characteristics. Dual y-axes depict the inversion time (black, left axis) and maximum $\Delta T$ (red, right axis). The results demonstrate that temperature-dependent parameters significantly affect the trends in inversion time and maximum $\Delta T$, with a nonmonotonic behavior observed in cases (a) and (b) due to the interplay between energy storage and dissipation. Further methodological details are provided in Section 3D Bowtie nanostructure of the main text.}
\end{figure}

\subsection*{Material Data Used in the Model}

Tables \ref{tab:thermal_properties} and \ref{tab:electronic_properties} summarize the material properties used in the model.

\begin{table}[t]
\centering
\caption{Thermal Properties of Materials}
\label{tab:thermal_properties}
\begin{tabular}{@{}llll@{}}
\toprule
\textbf{Material} & \textbf{Density} \(\rho\) 
                  & \textbf{Thermal Conductivity} \(\kappa\) 
                  & \textbf{Specific Heat} \(c\) \\
  & \([\,\mathrm{kg}\,\mathrm{m}^{-3}]\) 
  & \([\,\mathrm{W}\,\mathrm{m}^{-1}\,\mathrm{K}^{-1}]\) 
  & \([\,\mathrm{J}\,\mathrm{kg}^{-1}\,\mathrm{K}^{-1}]\) \\
\midrule
Gold (Au)      & 19300 & 350   & 128 \\
Platinum (Pt)  & 21400 & 77.8  & 125.06 \\
Nickel (Ni)    & 8900  & 90.9  & 440 \\
Copper (Cu)    & 8960  & 385   & 385 \\
Silver (Ag)    & 10500 & 429   & 235 \\
Aluminum (Al)  & 2700  & 237   & 897 \\
\bottomrule
\end{tabular}
\end{table}

\begin{table}[t]
\centering
\caption{Electronic Properties of Materials}
\label{tab:electronic_properties}
\begin{tabular}{@{}lll@{}}
\toprule
\textbf{Material} & \textbf{Fermi Energy} \(E_F\) & \textbf{Debye Temperature} \(T_D\)\\
  & \([\,\mathrm{eV}]\) & \([\,\mathrm{K}]\) \\
\midrule
Gold (Au)      & 5.53 & 170 \\
Platinum (Pt)  & 9.47 & 240 \\
Nickel (Ni)    & 5.15 & 450 \\
Copper (Cu)    & 7.00 & 343 \\
Silver (Ag)    & 5.48 & 225 \\
Aluminum (Al)  & 11.7 & 428 \\
\bottomrule
\end{tabular}
\end{table}

\subsection*{Discussion of Relevance}
The parameters presented in Tables \ref{tab:thermal_properties} and \ref{tab:electronic_properties} ensure general applicability across experimental setups. Specific points of relevance are as follows:

1. **Thermal Properties (\(\rho\), \(\kappa\), \(c\)):**
   - The thermal conductivity approximation, \(\kappa_e = \kappa_p \cdot (T_e / T_p)\), ensures a temperature-dependent behavior that aligns with observed trends in real materials.

2. **Electron-Phonon Coupling Coefficient and Heat Capacity:**
   - The electron-phonon coupling coefficient and the electronic heat capacity (\(C_e\)) are obtained from the open-access database at \href{https://compmat.org/electron-phonon-coupling/}{CompMat}.

3. **Refractive Index Data:**
   - The refractive index data for various materials, used in the model's optical calculations, are taken from \href{https://refractiveindex.info/}{RefractiveIndex.info}. Accurate refractive indices are essential for simulating optical absorption, which directly impacts heat generation and subsequent energy transfer between electrons and phonons.

These references and approximations form the basis of the model, ensuring its relevance for studying ultrafast dynamics and material responses under laser excitation.

\subsection*{Approximated 0D Model - Analytical solution}

Under the assumptions (i)-(iii) listed in the main text, the 0D model simplifies to:

\begin{equation}
\begin{cases}
$$C\dot{T_e}(t)=-G_{e-p}\Delta T_{e-p}(t)-\kappa_e T_e(t)+Q_0\tau \sqrt{2\pi}\delta(t-t_0)\\
C\dot{T_p}(t)=G_{e-p}\Delta T_{e-p}(t)$$
\end{cases}
\end{equation}

The analytical solution of the temperature inversion $\Delta T_{p-e}(t)$,  the time of the maximum inversion $t_{max}$ and the temperature inversion at its maximum, $\Delta T_{p-e}(t_{max})$ are:

\begin{equation}
\begin{cases}
$$\Delta T_{p-e}(t)=-\frac{e^{\frac{-tB}{2C}}\sqrt{\frac{\pi}{2}}Q_0(2G_{e-p}+\kappa_e+A+De^{\frac{tA}{C}})\tau}{CA}\\

t_{max}=\frac{2C}{A}\ln({\frac{2G_{e-p}+\kappa_e+A}{2G+\kappa_e-A})}\\

\Delta T_{p-e}(t_{max})=\frac{2B}{D'}(\frac{D'}{B})^\frac{B}{A}\frac{\sqrt{\frac{\pi}{2}}Q_0\tau}{C}$$
\end{cases}
\end{equation}

where $A=\sqrt{4G_{e-p}^2+\kappa_e^2}$, $B=2G_{e-p}+\kappa_e+A$ and $D=A-2G_{e-p}-\kappa_e$.

\section*{References}
\begin{itemize}
    \item[1] Computational Materials Database: Electron-Phonon Coupling,\\
    \href{https://compmat.org/electron-phonon-coupling/}{https://compmat.org/electron-phonon-coupling/}. Accessed on December 19, 2024.
    
    \item[2] Refractive Index Database,\\
    \href{https://refractiveindex.info/}{https://refractiveindex.info/}. Accessed on December 19, 2024.
\end{itemize}

\end{document}